\documentclass[journal]{IEEEtran}

\usepackage{latexsym}
\usepackage{graphicx}
\usepackage{amsfonts,amssymb,amsmath}
\usepackage{mathtools, cuted}
\usepackage{hyperref}

\usepackage[T1]{fontenc}
\usepackage{cite}
\usepackage{subcaption}
\usepackage{comment}

\usepackage{amsthm}
\usepackage{overpic}
\usepackage{steinmetz}
\usepackage{array}
\usepackage{url}
\usepackage{color}

\usepackage{algorithm}
\usepackage{algorithmic}

\usepackage{xcolor}
\usepackage{soul}

\usepackage{flushend}

\theoremstyle{plain}

\newtheorem{definition}{Definition}

\newtheorem{proposition}{Proposition}

\newcommand{\vect}[1]{\mathbf{#1}}

\def\diag{\mathrm{diag}}

\def\Htran{\mbox{\tiny $\mathrm{H}$}}
\def\Ttran{\mbox{\tiny $\mathrm{T}$}}
\def\CN{\mathcal{N}_{\mathbb{C}}} %Complex Gaussian
\begin{document}

\title{Dual-Polarized Reconfigurable Intelligent Surface Assisted Broad Beamforming}

\author{Parisa Ramezani, Maksym A. Girnyk, and Emil Bj\"{o}rnson

\thanks{%
P. Ramezani and E. Bj\"{o}rnson
are with the Department of Computer Science, KTH Royal Institute of Technology, Stockholm, Sweden (email:\{parram,emilbjo\}@kth.se). They are supported by Grant 2019-05068 from the Swedish Research Council. M. A. Girnyk is with Ericsson Networks, Stockholm, Sweden (e-mail: max.girnyk@ericsson.com).}}

\maketitle

\begin{abstract}

A reconfigurable intelligent surface (RIS) consists of a large number of low-cost elements that can control the propagation environment seen from a transmitter by intelligently applying phase shifts to impinging signals before reflection. This paper studies an RIS-assisted communication system where a transmitter wants to transmit a common signal to many users residing in a wide angular area. To cover this sector uniformly, the RIS needs to radiate a broad beam with a spatially flat array factor, instead of a narrow beam as normally considered. To achieve this, we propose to use a dual-polarized RIS consisting of elements with orthogonal polarizations and show that the RIS can produce a broad beam if the phase shift configuration vectors in the two polarizations form a so-called Golay complementary sequence pair. By utilizing their properties, we also present a method for constructing configuration for large RISs from smaller ones, while preserving the broad radiation pattern of the smaller RIS. The numerical results corroborate the mathematical analyses and highlight the greatly improved coverage properties.
\end{abstract}

\vspace{-1.5mm}


\vspace{-1.5mm}

\section{Introduction}

A reconfigurable intelligent surface (RIS) is a software-controlled meta-surface consisting of many passive elements (meta-atoms) that induce controllable phase shifts to the impinging signal to reflect it in a desired way \cite{Emil2019}. Since its emergence, RIS has been extensively studied to improve the performance of wireless networks in various ways, e.g., coverage extension, rank improvement, physical-layer security enhancement. 

The prior research on RIS-aided communication has focused on improving the performance of individual users by configuring the RIS to maximize their rate via forming narrow beams towards them during data transmission \cite{Guo2020,Li2020}. Meanwhile, there are many practical situations where a common signal must be transmitted to multitudes of users located in a wide angular range and possibly at unknown locations, e.g., when the base station transmits cell-specific signals such as Primary Synchronization Signal (PSS), Secondary Synchronization Signal (SSS) or Cell-specific Reference Signal (CRS) via the RIS. In such cases, a beam sweeping approach may be taken with the RIS forming narrow beams towards all possible angular directions, a technique that is not efficient in terms of resource utilization. An alternative is to let RIS reflect the signal as a broad beam to cover the large angular interval uniformly. However, the beamwidth is inversely proportional to the RIS aperture size, hence larger RISs produce narrower beams. Recently, the authors of \cite{He2023BroadCoverage} proposed a method for achieving a quasi-static broad coverage in RIS-assisted systems where they minimized the difference between a predefined pattern and the RIS-reflected power pattern by optimizing the base station precoder and RIS phase shifts. Their proposed scheme can partially widen the beam produced by the RIS which, however, exhibits fluctuations even in the limited range of target angles. This letter aims to devise a method for producing uniformly broad beams from the RIS which can cover all angular directions around the surface. 

A perfectly broad beam cannot be achieved in uni-polarized transmission unless only one single RIS element is activated.  
Recently, dual-polarized beamforming was utilized in \cite{Li2021Golay,Max2021,Max2022} to exploit the polarization degree of freedom of modern antennas to design arbitrarily broad beams when transmitting from a dual-polarized array. The principal idea is to design the beamforming weights of the dual-polarized antennas in a way that the beams corresponding to different polarizations complement each other, yielding an overall broad radiated beam.  Inspired by these works, this letter proposes the use of dual-polarized RIS composed of reflective elements with two orthogonal polarizations and proposes a phase shift design that enables broad beamforming from the RIS. 
 The contributions of this work can be summarized as follows: 1) We design the phase shifts of a dual-polarized RIS such that a broad beam is reflected from the RIS covering all possible angular directions where the users might reside. This is achieved by letting the RIS phase shift vectors be a so-called \textit{Golay complementary sequence pair}, having the unique property that their auto-correlation functions (ACFs) add up to a Kronecker delta function. 2) We show that this class of sequence pairs can be utilized to construct configurations for a large RIS from smaller ones while maintaining the beam shape. 3) We validate the analytical results numerically and show the effectiveness of the proposed scheme in providing good coverage and outperforming an RIS configured for diffuse scattering. 

The description of the system model is provided in Section \ref{sec:SysMod}. In Section \ref{sec:BroadBeamCondition}, we present the condition on the dual-polarized RIS configurations that must be met for producing a broad beam at the RIS. To meet the mentioned condition, we design the RIS configuration vectors in Section \ref{sec:RISDesign} by exploiting the unique properties of Golay complementary pairs and demonstrate how this class of complementary pairs can be used as radiation pattern-preserving expanders for constructing large RIS configuration vectors from smaller ones. Numerical simulations for verifying the mathematical analyses are provided in Section \ref{sec:NumRes}, and Section \ref{sec:Conclusions} concludes the paper.
\begin{figure}[t]
    \centering
    \includegraphics[width = 5cm]{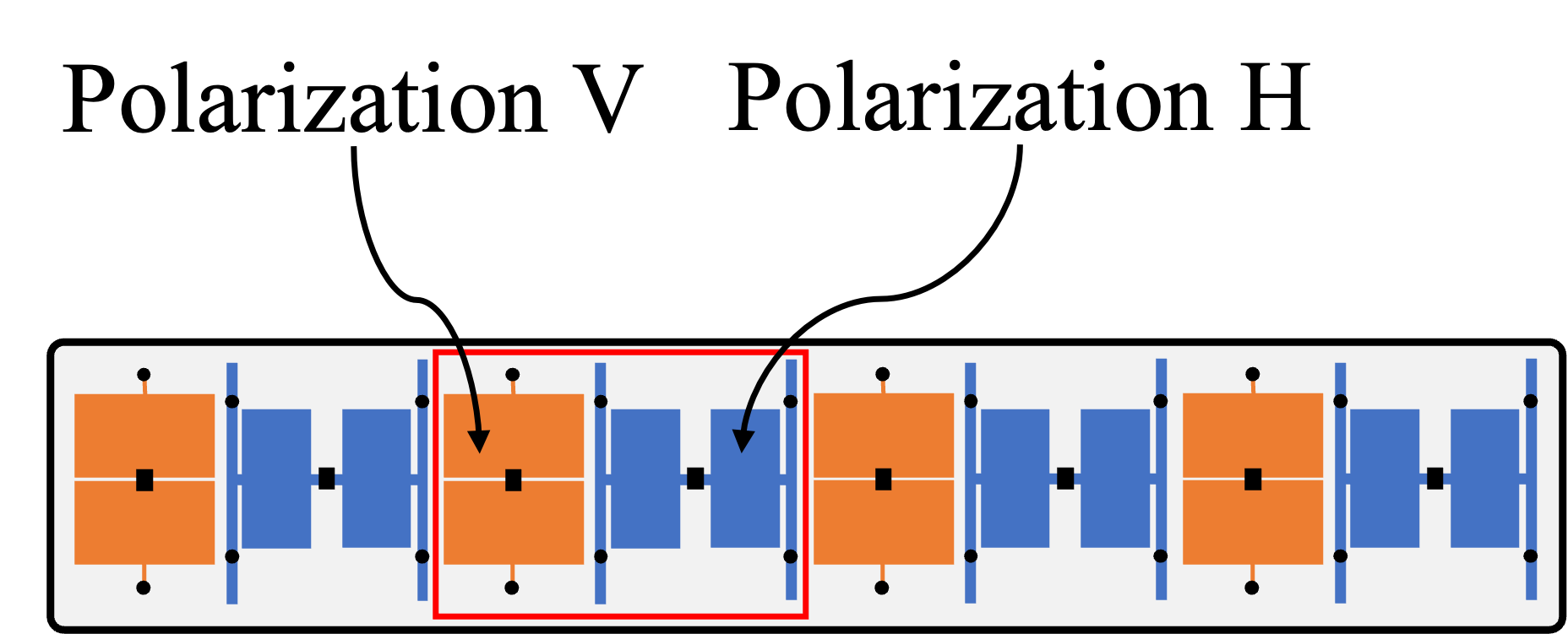}
    \vspace{0.1cm}
   \caption{A dual-polarized RIS where each meta-atom consists of two reflectors that reflect signals with orthogonal polarizations and thereby can independently modify the phase of the impinging waves.  See \cite{ke2021linear} for practical implementation details.} 
    \label{fig:DP-RIS}
    \vspace{-4mm}
\end{figure}
\vspace{-3mm}
\section{System Model}
\label{sec:SysMod}
We consider a downlink RIS-assisted communication system where a transmitter intends to transmit a common signal to multiple users through an RIS. The transmitter and the users are equipped with one dual-polarized antenna each, while the RIS has $2M$ elements with a horizontal uniform linear array (ULA) topology. Out of the $2M$ RIS elements, there are $M$ elements with polarization H and $M$ elements with polarization V.\footnote{H and V refer to horizontal and vertical polarizations, but the results of this letter are applicable to any pair of orthogonal polarizations.} An example of a dual-polarized ULA RIS structure is illustrated in Fig.~\ref{fig:DP-RIS}. 

Suppose that $s \sim \CN{(0,1)}$ is the transmitted signal. In a line-of-sight (LoS) scenario,  the received signal at an arbitrary user in polarization $i$ can be expressed as\footnote{We assume a LoS channel between the transmitter and RIS, which is the most practical scenario \cite{Emil2022}. LoS between the RIS and receiver is assumed because we want to achieve uniform array gain in all LoS directions.}
\begin{equation}
\label{eq:received_signal}
r_i = \sqrt{P}\vect{g}_i^T \boldsymbol{\Phi}_i \vect{h}_i s + n_i,~~i \in \{\mathrm{H},\mathrm{V}\}, 
\end{equation}
where $P$ is the transmit power, and $\vect{h}_i \in \mathbb{C}^M$ and $\vect{g}_i \in \mathbb{C}^{M}$ represent the transmitter-RIS and RIS-user channels, respectively, in polarization $i$. Further, $\boldsymbol{\Phi}_i$ is the phase shift configuration matrix of the RIS in this polarization with $\boldsymbol{\Phi}_i = \diag{\left(e^{j\phi_{i,1}},\ldots,e^{j\phi_{i,M}}\right)}$ where $\phi_{i,m}$ denotes the phase shift applied by the $m$th RIS element in polarization $i$ to the incident signal. Finally, $n_i \sim \CN{(0,\sigma^2)}$ is the receiver noise. 
The LoS channels can be expressed as $\vect{h}_i =  \sqrt{\beta_hG_0(\Tilde{\varphi})}\vect{a}_i (\Tilde{\varphi})$ and $\vect{g}_i =  \sqrt{\beta_g G_0(\varphi)}\vect{a}_i(\varphi)$, where $\beta_h$ and $\beta_g$ denote the path-loss at the reference RIS element. $\Tilde{\varphi}$ denotes the angle of arrival (AoA) from the transmitter to the RIS, which is assumed to be known and fixed at the deployment stage, $\varphi$ is the angle of departure (AoD) from the RIS to the user, $G_0(\cdot)$ is the radiation power pattern of a single RIS element, and $\vect{a}_i(\cdot)$ is the RIS array response vector for polarization $i$. 
The array response vectors can be expressed as \cite{massivemimobook} 
\begin{align}
    \!\!\!\vect{a}_{\mathrm{H}} (x) \!&=\!  [1, e^{-j \frac{2\pi}{\lambda} (2\Delta_d) \sin(x)}\!, \ldots, e^{-j(M -1)\frac{2\pi}{\lambda} (2\Delta_d) \sin(x)} ]^{\Ttran}\!\!\!, \!\\
    \!\!\!\vect{a}_{\mathrm{V}} (x) \!&= \!e^{-j\frac{2\pi}{\lambda} \Delta_d \sin(x)} \vect{a}_{\mathrm{H}} (x),
\end{align}
where $\Delta_d$ is the inter-element spacing between adjacent RIS elements and $\lambda$ is the wavelength of the transmitted signal. 
Setting $\boldsymbol{\phi}_i = \mathrm{diag}(\boldsymbol{\Phi}_i)$, we can rewrite
\eqref{eq:received_signal} as 
\begin{equation}
\label{eq:received-signal2}
    r_i = \sqrt{PG_0(\Tilde{\varphi})G_0(\varphi)\beta_h\beta_g}  \boldsymbol{\phi}_i^{\Ttran} \big(\vect{a}_i(\Tilde{\varphi}) \odot \vect{a}_i (\varphi)\big)s + n_i
\end{equation}
and apply maximum ratio combining over the two polarizations to obtain the signal-to-noise ratio (SNR) at the user as
\begin{align}
\label{eq:SNR}
   &\mathrm{SNR} = \frac{P\beta_h \beta_g}{\sigma^2} G_0(\Tilde{\varphi})G_0(\varphi) A(\varphi), 
\end{align}where the term 
\begin{equation}
\label{eq:array-factor}
    A(\varphi) \!=\! \Big|\boldsymbol{\phi}_{\mathrm{H}}^{\Ttran} \big( \vect{a}_{\mathrm{H}}(\Tilde{\varphi}) \odot \vect{a}_{\mathrm{H}}(\varphi)\big)\Big|^2 \!\!+ \Big|\boldsymbol{\phi}_{\mathrm{V}}^{\Ttran} \big( \vect{a}_{\mathrm{V}}(\Tilde{\varphi}) \odot \vect{a}_{\mathrm{V}}(\varphi)\big)\Big|^2 
\end{equation} 
is called the \textit{power-domain array factor} (PDAF) of the dual-polarized RIS.  The total radiation power pattern observed by the user is therefore given by $G(\varphi) = A(\varphi) \hat{G_0}(\varphi)$ with $\hat{G_0}(\varphi) = G_0(\Tilde{\varphi}) G_0(\varphi)$. The aim of this paper is to design the PDAF such that the signal radiated from the RIS forms a broad beam covering the entire angular interval where the prospective users might reside. The proposed procedure is elaborated in the next section.

\section{Broad Beam Condition}
\label{sec:BroadBeamCondition}
To radiate a maximally broad beam from the RIS, the PDAF \emph{must be} 
spatially flat, i.e., it must be constant over all azimuth angles in front of the RIS:
\begin{equation}
\label{eq:AF-constant}
\Big|\boldsymbol{\phi}_{\mathrm{H}}^{\Ttran} \hat{\vect{a}}_{\mathrm{H}}(\Tilde{\varphi},\varphi)\Big|^2 + \Big|\boldsymbol{\phi}_{\mathrm{V}}^{\Ttran} \hat{\vect{a}}_{\mathrm{V}}(\Tilde{\varphi},\varphi)\Big|^2=c, ~ \varphi \in \left[-\frac{\pi}{2},\frac{\pi}{2} \right], 
\end{equation}
where $c$ is a constant and 
\begin{align}
   \!\!\!\!\hat{\vect{a}}_{\mathrm{H}} (\Tilde{\varphi},\varphi) \!&=\! \vect{a}_{\mathrm{H}}(\Tilde{\varphi}) \odot \vect{a}_{\mathrm{H}}(\varphi) \! =\! [1,e^{-j2\psi}\!\!,\ldots,e^{-j2(M-1)\psi}]^{\Ttran}\!\!\!, \\
   \!\!\!\!\hat{\vect{a}}_{\mathrm{V}} (\Tilde{\varphi},\varphi) \!&=\! \vect{a}_{\mathrm{V}}(\Tilde{\varphi}) \odot \vect{a}_{\mathrm{V}}(\varphi) \! =\! e^{-j\psi} \hat{\vect{a}}_{\mathrm{H}}(\Tilde{\varphi},\varphi)
\end{align}
are the equivalent array response vectors for the H and V polarizations, respectively, where the relative phase shift reads as
\begin{equation}
\label{eq:psi}
      \psi = \frac{2\pi}{\lambda}\Delta_d \big( \sin(\Tilde{\varphi}) + \sin (\varphi)\big).
\end{equation}
By expanding the phase shift and array response vectors, the condition in \eqref{eq:AF-constant} can be rewritten as 
\begin{equation}
\label{eq:AF-constant2}
 \!\!\!\Big| \!\sum_{m=1}^M [\boldsymbol{\phi}_\mathrm{H}]_m e^{-j2(m-1)\psi} \Big|^2 \!+  \Big| \!\sum_{m=1}^M [\boldsymbol{\phi}_\mathrm{V}]_m e^{-j(2m-1)\psi} \Big|^2 \!\!= c,\!
\end{equation}
where $[\boldsymbol{\phi}_i]_m$ represents the $m$th entry of the vector $\boldsymbol{\phi}_i$. The first and second terms in \eqref{eq:AF-constant2} are the square of the (spatial) discrete-time Fourier transform of the sequences $\boldsymbol{\phi}_{\mathrm{H}}$ and $\boldsymbol{\phi}_{\mathrm{V}}$, respectively, if these are padded with zero.   
Since the power spectral density (PSD) of a sequence equals the square of its Fourier transform \cite[Chapter 4]{Stein2000}, \eqref{eq:AF-constant2}
can be expressed as
\begin{equation}
\label{eq:sum_spectral_density}
 S_{\boldsymbol{\phi}_{\mathrm{H}}}(\psi) + S_{\boldsymbol{\phi}_{\mathrm{V}}}(\psi) = c,
\end{equation}
where $S_{\boldsymbol{\phi}}(\psi)$ denotes the PSD of $\boldsymbol{\phi}$. According to the Wiener-Khinchin theorem, the PSD and ACF are Fourier transform pairs \cite{Wiener1930}. Hence, taking the Fourier transform of both sides of \eqref{eq:sum_spectral_density}, the requirement of a spatially flat PDAF translates into  
\begin{equation}
\label{eq:sum_ACF}
   R_{\boldsymbol{\phi}_{\mathrm{H}}}[\tau] + R_{\boldsymbol{\phi}_{\mathrm{V}}}[\tau]  = c\delta [\tau],
\end{equation}
where $\delta[\tau]$ is the Kronecker delta factunion and $R_{\boldsymbol{\phi}}[\tau]$ indicates the ACF of the vector $\boldsymbol{\phi} \in \mathbb{C}^M$, which is given by 

 \begin{align}
    \label{eq:ACF}
    	R_{\boldsymbol{\phi}}[\tau] = \left\{ 
    	\begin{matrix} 
    	&\sum\limits_{m=1}^{M-\tau} [\boldsymbol{\phi}]_m [\boldsymbol{\phi}]_{m+\tau}^*, 
    	& & \tau=0,\ldots,M-1,\\
    	&\sum\limits_{m=1}^{M+\tau} [\boldsymbol{\phi}]_{m-\tau} [\boldsymbol{\phi}]_{m}^*, 
    	& &\tau = -M+1,\ldots,-1,\\
    	& 0,
    	& & \mathrm{otherwise}
    	\end{matrix} \right. 
    \end{align}
    The equality in \eqref{eq:sum_ACF} indicates that the sum of the ACFs of $\boldsymbol{\phi}_{\mathrm{H}}$ and $\boldsymbol{\phi}_{\mathrm{V}}$ must be zero except at $\tau=0$, where it becomes
\begin{equation}
R_{\boldsymbol{\phi}_{\mathrm{H}}} [0] +  R_{\boldsymbol{\phi}_{\mathrm{V}}} [0] =  \sum_{m=1}^M \Big(\Big|[\boldsymbol{\phi}_{\mathrm{H}}]_m \Big|^2 + \Big|[\boldsymbol{\phi}_{\mathrm{V}}]_m \Big|^2\Big) = 2M,
\end{equation} 
from which we can identify $c = 2M$. 
Hence, \eqref{eq:sum_ACF} turns into
\begin{equation}
\label{eq:broad-beam-condition}
   R_{\boldsymbol{\phi}_{\mathrm{H}}}[\tau] + R_{\boldsymbol{\phi}_{\mathrm{V}}}[\tau] = 2 M\delta [\tau].
\end{equation}
We refer to \eqref{eq:broad-beam-condition} as the \textit{broad beam condition} for RIS-aided communication because the configuration pair $(\boldsymbol{\phi}_{\mathrm{H}},\boldsymbol{\phi}_{\mathrm{V}})$ must satisfy \eqref{eq:broad-beam-condition} for the RIS to radiate a maximally broad beam. 

\section{RIS Phase Configuration Design}
\label{sec:RISDesign}
The objective is to design the RIS phase configuration vectors $(\boldsymbol{\phi}_{\mathrm{H}},\boldsymbol{\phi}_{\mathrm{V}})$ such that the PDAF of the RIS becomes a constant for all observation angles. To this end, we first describe a class of sequence pairs called \textit{Golay complementary sequence pairs}, introduced by Golay in \cite{Golay1951}. 

\begin{definition}[Golay sequence pair]
\label{def:Golay-pair}
    Unimodular sequences $\vect{u} \in \mathbb{C}^M$ and $\vect{v} \in \mathbb{C}^M$ form a Golay complementary pair if 
    \begin{align}
    \label{eqn:conditionGolay}
    	R_{\vect{u}}[\tau] + R_{\vect{v}}[\tau] &= 2M \delta [\tau].
    \end{align}
\end{definition} 

From this definition, a pair of Golay complementary sequences possess the unique property that their ACFs add up to a Kronecker delta. This results in the following proposition.

\begin{proposition}
  A dual-polarized RIS radiates a broad beam if its phase configuration vectors form a Golay pair.
\end{proposition}

\begin{figure}
	\centering
	\begin{subfigure}{0.495\textwidth}
		\centering
		\includegraphics[width=\columnwidth]{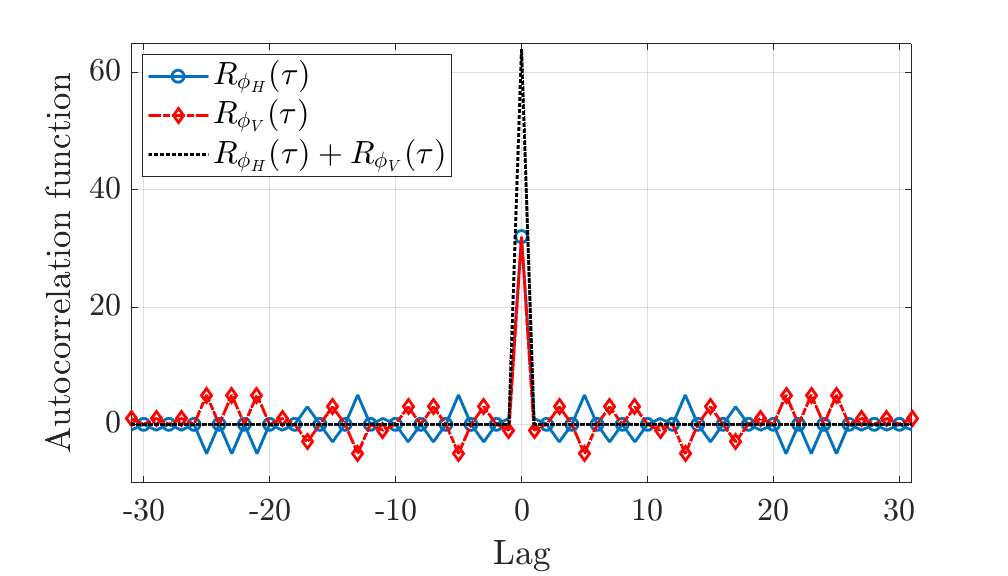}
		\caption{Auto-correlation function.}
		\label{fig:Golay-ACF}
	\end{subfigure}
	\begin{subfigure}{0.495\textwidth}
		\centering
		\includegraphics[width=\columnwidth]{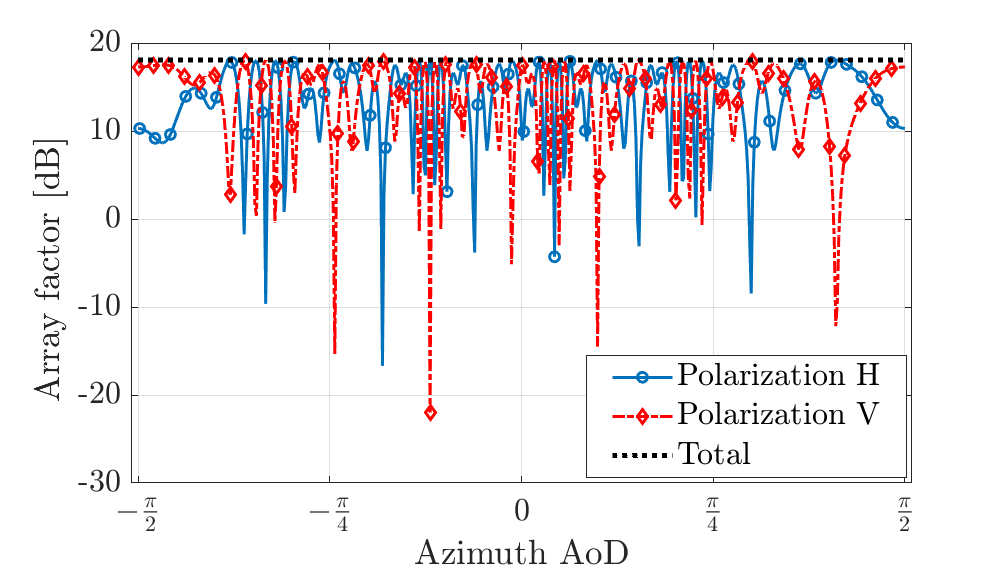}
		\caption{Power-domain array factor.}
		\label{fig:Golay-AF}
	\end{subfigure}
	\caption{Illustration of the sum ACF and PDAF for a Golay complementary pair of length $M=32$.}
 \vspace{-6mm}
\end{figure}

\begin{IEEEproof}
   Let $(\boldsymbol{\phi}_{\mathrm{H}},\boldsymbol{\phi}_{\mathrm{V}})$ form a Golay complementary pair of length $M$. From Definition \ref{def:Golay-pair}, we have 
\begin{align}
    R_{\boldsymbol{\phi}_{\mathrm{H}}}[\tau] + R_{\boldsymbol{\phi}_{\mathrm{V}}}[\tau] = 2 M\delta [\tau].
\end{align}Therefore,  $(\boldsymbol{\phi}_{\mathrm{H}},\boldsymbol{\phi}_{\mathrm{V}})$ satisfy the broad beam condition in \eqref{eq:broad-beam-condition} and a $2M$-element dual-polarized RIS with $(\boldsymbol{\phi}_{\mathrm{H}},\boldsymbol{\phi}_{\mathrm{V}})$ as its configuration pair radiates a broad beam.  
\end{IEEEproof}
Fig. \ref{fig:Golay-ACF} illustrates an example of the sum ACF of an RIS configuration pair $(\boldsymbol{\phi}_{\mathrm{H}},\boldsymbol{\phi}_{\mathrm{V}})$ of length $M=32$ which form a Golay complementary pair. It can be seen that the sum ACF is zero except at $\tau = 0$ where we have $R_{\boldsymbol{\phi}_{\mathrm{H}}}[0] + R_{\boldsymbol{\phi}_{\mathrm{V}}}[0] = 2M = 64$. Fig. \ref{fig:Golay-AF} illustrates the PDAF for the RIS with the same configuration vectors. We can observe that the total array factor is constant over all observation angles with its value being given by $10\log_{10}(64) \approx 18$.

Golay complementary pairs can be used to expand a small RIS to another RIS with larger number of elements while preserving the radiation pattern of the former. Specifically, using Golay pairs, the phase shift configuration vectors of the primary RIS can be expanded to phase shift configuration vectors of a larger size without affecting the shape of the beam radiated by the primary RIS. The following proposition elaborates on how Golay complementary pairs act as expanders for creating larger configuration vectors from smaller ones. 

\begin{proposition}
\label{prop:Golay-expansion}
Assume that $\boldsymbol{\phi}_{\mathrm{H}} $ and $\boldsymbol{\phi}_{\mathrm{V}}$ are the configuration vectors of an RIS with $2M$ elements with $A(\varphi)$ as its PDAF. Further, let $(\vect{u}, \vect{v})$ be a Golay pair of length $N$. Then, an expanded RIS with $4MN$ elements and configuration vectors
\begin{align}
    	\label{eq:Golay-expansion}
    	\Tilde{\boldsymbol{\phi}}_{\mathrm{H}} = \begin{bmatrix}
    		\vect{u} \otimes \boldsymbol{\phi}_{\mathrm{H}}\\
    		-\vect{v} \otimes \vect{E}_M \boldsymbol{\phi}_{\mathrm{V}}^*
    	\end{bmatrix},~~~
    	\Tilde{\boldsymbol{\phi}}_{\mathrm{V}} = \begin{bmatrix}
    		\vect{u} \otimes \boldsymbol{\phi}_{\mathrm{V}}\\
    		\vect{v} \otimes \vect{E}_M \boldsymbol{\phi}_{\mathrm{H}}^*
    	\end{bmatrix}\!,
    \end{align}preserves the radiation pattern of the primary RIS. Specifically,  the PDAF of the expanded RIS becomes
    \begin{equation}
        \Tilde{A}(\varphi) = 2N A(\varphi).
    \end{equation} 
\end{proposition}

\begin{IEEEproof} The proof is given in the appendix. 
\end{IEEEproof}

\section{Numerical Results}
\label{sec:NumRes}
In this section, we present our numerical results. Fig.~\ref{fig:Golay-Expansion} shows how the design for a primary RIS with $20$ elements can be expanded to an RIS with $120$ elements while preserving the radiation power pattern of the primary RIS. The element spacing is set as $\Delta_d  = \lambda/2$, the AoA from the transmitter to the RIS is chosen to be $\Tilde{\varphi} = \pi/3$, and the radiation pattern of each element is obtained using the 3GPP model \cite{3gpp}:
\begin{equation}
    G_0(\varphi) = 8 -\min \Big(12 \big( \frac{\varphi - \varphi_0}{\Delta \varphi}\big)^2,30 \Big) \,\, \textrm{[dBi]},
\end{equation}
with $\varphi_0 = 0$ and $\Delta \varphi = \pi/2$ \cite{tiwari2022ris}. The primary RIS with the total number of elements of $2M = 20$ is configured to produce a broad beam spanning the angular interval $\varphi \in [-\pi/2,~\pi/2]$ where its phase configuration vectors are chosen to be the following Golay complementary pairs from \cite{Holzmann94}:
\begin{figure}[t!]
\centering
	\begin{overpic}[width=\columnwidth,tics=10]{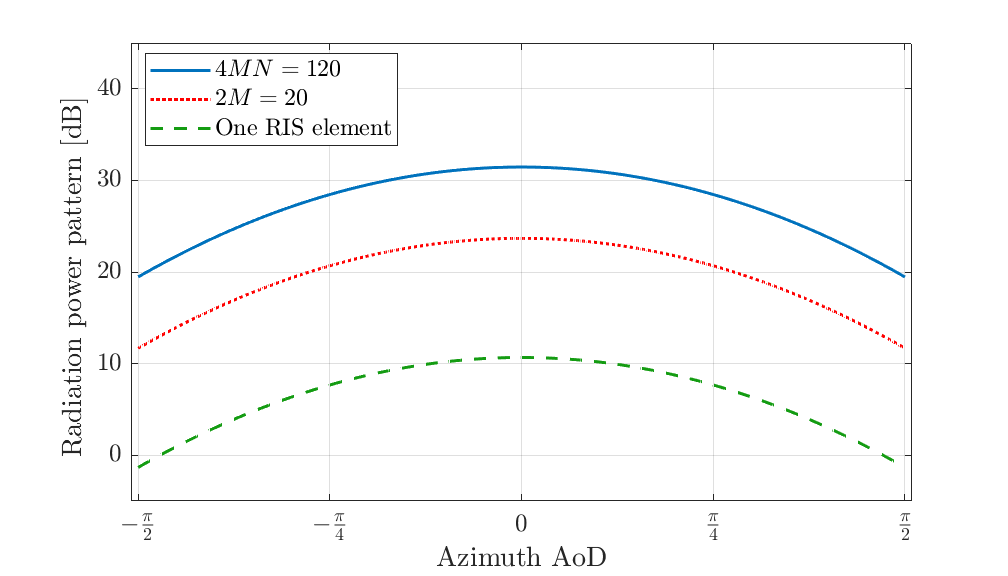}
	 \put(52,37.4){\vector(0,1){3.3}}
      \put(52,37.4){\vector(0,-1){3.3}}
      \put(53,36.1){\small $10\log2N$}
      \put(52,27.8){\vector(0,1){5.7}}
      \put(52,27.8){\vector(0,-1){5.7}}
      \put(53,26.5){\small $10\log2M$}
\end{overpic} 
\caption{Radiation power pattern vs. azimuth AoD.
}  \label{fig:Golay-Expansion}
\vspace{-4mm}
\end{figure}
\begin{figure}[t]
    \centering
    \includegraphics[width =\columnwidth]{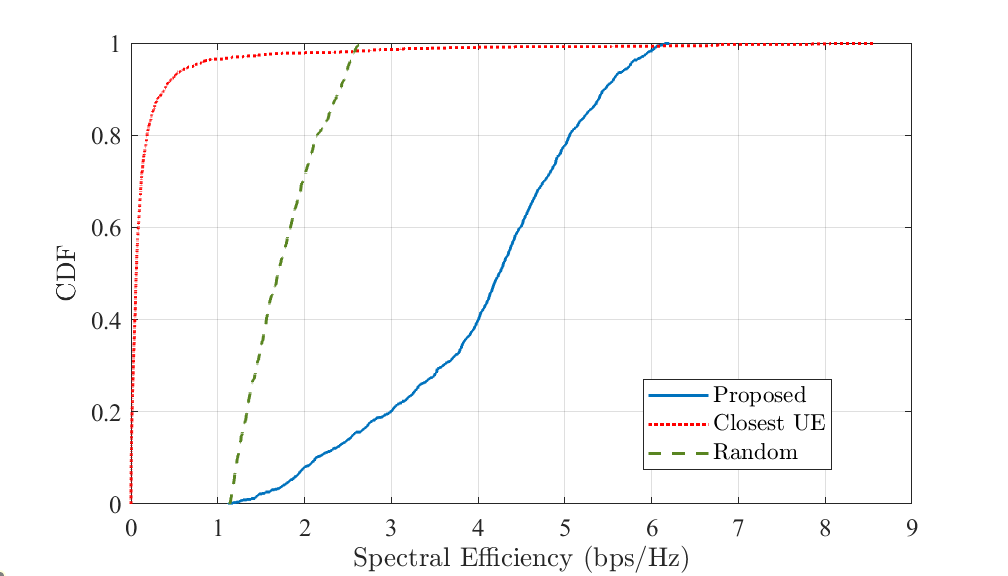}
   \caption{CDF of spectral efficiency.}
    \label{fig:Golay-CDF}
    \vspace{-4mm}
\end{figure}
\begin{align}
\label{eq:primary-config}
   \boldsymbol{\phi}_{\mathrm{H}} &= \exp \left(j\left[0,0,\pi,\frac{\pi}{2},0,\pi,-\frac{\pi}{2},\pi,-\frac{\pi}{2},\frac{\pi}{2}\right]\right), \notag \\
   \boldsymbol{\phi}_{\mathrm{V}} &= \exp \left(j\left[0,0,-\frac{\pi}{2},0,-\frac{\pi}{2},-\frac{\pi}{2},0,\frac{\pi}{2},\frac{\pi}{2},-\frac{\pi}{2}\right]\right) .
\end{align}
The dotted red line in Fig.~\ref{fig:Golay-Expansion} shows the radiation pattern of the primary RIS, which matches that of a single RIS element (dashed green line) except for being shifted upwards. The difference between these two curves is  $10\log_{10}(2M) \approx 13\,$dB for all azimuth AoDs which indicates the gain achieved by employing an RIS with $2M = 20$ elements. Hence, the PDAF is constant as expected from the theory. The solid blue curve shows the radiation pattern of an expanded RIS with $4MN = 120$ elements for which the phase configuration vectors are obtained based on Proposition \ref{prop:Golay-expansion}, where Golay pairs of length $N=3$ are used to form the configuration vectors of the expanded RIS. In particular, with the phase shift configuration vectors of the primary RIS given in \eqref{eq:primary-config} and the Golay complementary pairs selected based on \cite{Holzmann94} as
\begin{equation}
  \vect{u} =  \exp \left(j\left[0,\frac{\pi}{2},0\right]\right),~~ \vect{v} =  \exp \left(j\left[0,0,\pi\right]\right),
\end{equation}
the new phase shift configurations are obtained from \eqref{eq:Golay-expansion}. Fig.~\ref{fig:Golay-Expansion} demonstrates that this expansion shifts the radiation pattern of the primary RIS upwards by $10\log_{10}(2N) \approx 7.8$~dB. The difference between the received power with only one RIS element and with an RIS with $4MN = 120$ elements is $10\log(120) \approx 20.8\,$dB over all azimuth AoDs which once again demonstrates that our design preserves the broad radiation pattern of one element.  

Next, we consider an RIS-assisted downlink setting with many users residing in a specific geographical area. 
We perform a Monte Carlo simulation with $K=1000$ prospective user locations uniformly distributed over the coverage area. 
The spectral efficiency (SE) of the $k$th user ($\mathrm{U}_k$) is given by  
\begin{equation}
\label{eq:SE-expression}
    \mathrm{SE}_k = \log_2 \Big(1+\frac{P}{\sigma_k^2} \beta_h \beta_{g,k} G(\varphi_k) \Big)~~ \mathrm{[bps/Hz]},
\end{equation}where the path-loss is modeled as \cite{Emil2020}
\begin{equation}
    \beta_h (\beta_{g,k}) = -37.5 - 22 \log \big(d_h (d_{g,k})/1~ \mathrm{m}\big) ~~[\mathrm{dB}],
\end{equation}
with $d_h$ and $d_{g,k}$ being the distances between the transmitter and the RIS and  between the RIS and $\mathrm{U}_k$, respectively. Moreover, $\varphi_k$ is the AoD from the RIS to $\mathrm{U}_k$. We set $d_h = 50$~m and assume that the users are uniformly distributed around the RIS, such that $d_{g,k} \sim \mathcal{U}[50~\mathrm{m},100~\mathrm{m}]$ and $\varphi_k \sim \mathcal{U}[-\pi/2,\pi/6]$. 
The RIS is assumed to have $M = 60$ elements per polarization. The transmit power is set as $P=47$~dBm and the noise power is given by $\sigma_k^2 = -90~\mathrm{dBm}, \forall k$.

In Fig.~\ref{fig:Golay-CDF}, the cumulative distribution function (CDF) of the SE delivered by the proposed scheme to different parts of the coverage area is compared with two benchmarks. The first benchmark represents the scenario where the RIS phase shifts are chosen to maximize the SE of the user with the best channel condition to the RIS, which will be the user closest to it in the broadside direction. The RIS phase shifts in each polarization then matches with the DFT vector determined by the closest user's angle, resulting in a narrow beam that maximizes that user's received power.
This scheme is labeled as ``Closest UE'' in the figure. The second benchmark corresponds to the case where the RIS elements act as diffuse scatterers by applying random phase shifts to the incident signal before reflecting it. This case is tagged as ``Random'' in the figure. With the “Closest UE” scheme, more than $96\%$ of the users experience less than $1\,$bps/Hz of SE. Although this scheme can provide a few users with a SE above $8\,$bps/Hz, most of the users are poorly served because the RIS produces a narrow beam towards a specific user location. Our proposed approach can effectively provide an acceptable beamforming gain in all the desired directions simultaneously, as a result of which $93 \%$ of the users achieve more than $2$ bps/Hz. 
The random scattering scheme performs much better than the “Closest UE” scheme in terms of fairness but cannot deliver more than $2.64\,$bps/Hz SE due to poor configuration of RIS elements in both polarizations. The proposed broad beamforming design is able to uniformly cover the wide angular area of interest with an acceptable level of SE.  

\section{Conclusions}
\label{sec:Conclusions}
This paper investigated an RIS-aided communication scenario with a transmitter sending a common signal to a large number of users with the assistance of a dual-polarized RIS. We proposed a novel broad-beam design characterized by a spatially flat PDAF so that the RIS becomes equally helpful for all users located in any angular direction.
Specifically, we proved that to achieve a spatially flat PDAF, the RIS phase shifts in the two polarizations must be co-designed so that their ACFs add up to a Kronecker delta. This unique property is possessed by a class of RIS configurations called Golay complementary sequence pairs. Thus, by letting RIS phase shift vectors form a Golay pair, the RIS can radiate a broad beam and simultaneously serve a large number of users. We also proposed a method for RIS expansion using Golay pairs, where the configuration designed for a small RIS is expanded to be used with a larger RIS while retaining its beam shape.

\appendix

To prove Proposition~\ref{prop:Golay-expansion}, consider the expanded RIS' PDAF:
\begin{equation}
\label{eq:AF-expanded}
    \Tilde{A}(\varphi) = \big|\Tilde{\boldsymbol{\phi}}_{\mathrm{H}}^{\Ttran}\hat{\vect{a}}_{\mathrm{H}}(\Tilde{\varphi},\varphi)\big|^2 + \big|\Tilde{\boldsymbol{\phi}}_{\mathrm{V}}^{\Ttran}\hat{\vect{a}}_{\mathrm{V}}(\Tilde{\varphi},\varphi)\big|^2.
\end{equation}For notational simplicity, we henceforth omit the argument $(\Tilde{\varphi},\varphi)$ from the equivalent array response vector expressions. Since $\hat{\vect{a}}_{\mathrm{V}}$ is a phase shifted version of $\hat{\vect{a}}_{\mathrm{H}}$, the former can be replaced by the latter in \eqref{eq:AF-expanded} without affecting $\Tilde{A}(\varphi)$.  The expanded RIS of size $4MN$ consists of two sub-RISs, each having $2MN$ elements with $MN$ elements per polarization. The RIS phase configuration vectors for polarizations H and V of the two sub-RISs are given by 
\begin{align}
    &\Tilde{\boldsymbol{\phi}}_{\mathrm{H},1} = \vect{u} \otimes \boldsymbol{\phi}_{\mathrm{H}},~~&&\Tilde{\boldsymbol{\phi}}_{\mathrm{V},1} = \vect{u} \otimes \boldsymbol{\phi}_{\mathrm{V}}, \\
    &\Tilde{\boldsymbol{\phi}}_{\mathrm{H},2} = -\vect{v} \otimes \vect{E}_M \boldsymbol{\phi}_{\mathrm{V}}^*,~&&\Tilde{\boldsymbol{\phi}}_{\mathrm{V},2} = \vect{v} \otimes \vect{E}_M \boldsymbol{\phi}_{\mathrm{H}}^*.
\end{align}The array response vector $\hat{\vect{a}}_{\mathrm{H}}$ can then be expressed as
\begin{equation}
    \hat{\vect{a}}_{\mathrm{H}} = [\hat{\vect{a}}_{\mathrm{H},1}^{\Ttran},\hat{\vect{a}}_{\mathrm{H},2}^{\Ttran}]^{\Ttran}  \in \mathbb{C}^{2MN},
\end{equation} 
where $\hat{\vect{a}}_{\mathrm{H},1} \in \mathbb{C}^{MN}$ and $\hat{\vect{a}}_{\mathrm{H},2} \in \mathbb{C}^{MN}$ are the equivalent array response vectors of the two sub-RISs. The PDAF of the expanded array is then obtained as 
\begin{align}
\label{eq:AF-expanded-2}
   & \Tilde{A} (\varphi) = \Big|\big[\Tilde{\boldsymbol{\phi}}_{\mathrm{H},1}^{\Ttran}, \Tilde{\boldsymbol{\phi}}_{\mathrm{H},2}^{\Ttran} \big] \begin{bmatrix}
    		\hat{\vect{a}}_{\mathrm{H},1}\\
    		\hat{\vect{a}}_{\mathrm{H},2}
    	\end{bmatrix} \Big|^2 + \Big|\big[\Tilde{\boldsymbol{\phi}}_{\mathrm{V},1}^{\Ttran}, \Tilde{\boldsymbol{\phi}}_{\mathrm{V},2}^{\Ttran} \big] \begin{bmatrix}
    		\hat{\vect{a}}_{\mathrm{H},1}\\
    		\hat{\vect{a}}_{\mathrm{H},2}
    	\end{bmatrix}  \Big|^2 \notag \\
     & = \big| \Tilde{\boldsymbol{\phi}}_{\mathrm{H},1}^{\Ttran} \hat{\vect{a}}_{\mathrm{H},1} + \Tilde{\boldsymbol{\phi}}_{\mathrm{H},2}^{\Ttran} \hat{\vect{a}}_{\mathrm{H},2}\big|^2 + \big| \Tilde{\boldsymbol{\phi}}_{\mathrm{V},1}^{\Ttran} \hat{\vect{a}}_{\mathrm{H},1} + \Tilde{\boldsymbol{\phi}}_{\mathrm{V},2}^{\Ttran} \hat{\vect{a}}_{\mathrm{H},2}\big|^2 \notag \\
     & = \big| \Tilde{\boldsymbol{\phi}}_{\mathrm{H},1}^{\Ttran} \hat{\vect{a}}_{\mathrm{H},1}\big|^2 + \big| \Tilde{\boldsymbol{\phi}}_{\mathrm{H},2}^{\Ttran} \hat{\vect{a}}_{\mathrm{H},2}\big|^2 + 2 \mathrm{Re}(\hat{\vect{a}}_{\mathrm{H},1}^{\Htran} \Tilde{\boldsymbol{\phi}}_{\mathrm{H},1}^* \Tilde{\boldsymbol{\phi}}_{\mathrm{H},2}^{\Ttran} \hat{\vect{a}}_{\mathrm{H},2}) \notag \\
     & + \big| \Tilde{\boldsymbol{\phi}}_{\mathrm{V},1}^{\Ttran} \hat{\vect{a}}_{\mathrm{H},1}\big|^2 + \big| \Tilde{\boldsymbol{\phi}}_{\mathrm{V},2}^{\Ttran} \hat{\vect{a}}_{\mathrm{H},2}\big|^2 + 2 \mathrm{Re}(\hat{\vect{a}}_{\mathrm{H},1}^{\Htran} \Tilde{\boldsymbol{\phi}}_{\mathrm{V},1}^* \Tilde{\boldsymbol{\phi}}_{\mathrm{V},2}^{\Ttran} \hat{\vect{a}}_{\mathrm{H},2}) \notag \\
     & =  \big| \Tilde{\boldsymbol{\phi}}_{\mathrm{H},1}^{\Ttran} \hat{\vect{a}}_{\mathrm{H},1}\big|^2 + \big| \Tilde{\boldsymbol{\phi}}_{\mathrm{H},2}^{\Ttran} \hat{\vect{a}}_{\mathrm{H},2}\big|^2 +  \big|\Tilde{\boldsymbol{\phi}}_{\mathrm{V},1}^{\Ttran} \hat{\vect{a}}_{\mathrm{H},1}\big|^2 + \big| \Tilde{\boldsymbol{\phi}}_{\mathrm{V},2}^{\Ttran} \hat{\vect{a}}_{\mathrm{H},2}\big|^2 \notag \\
     & + 2\mathrm{Re} \Big( \hat{\vect{a}}_{\mathrm{H},1}^{\Htran} \big(\Tilde{\boldsymbol{\phi}}_{\mathrm{H},1}^* \Tilde{\boldsymbol{\phi}}_{\mathrm{H},2}^{\Ttran}  + \Tilde{\boldsymbol{\phi}}_{\mathrm{V},1}^* \Tilde{\boldsymbol{\phi}}_{\mathrm{V},2}^{\Ttran} \big)\hat{\vect{a}}_{\mathrm{H},2}\Big).
\end{align} We first consider the last term in \eqref{eq:AF-expanded-2}. We have 
\begin{align}
  &  \hat{\vect{a}}_{\mathrm{H},1}^{\Htran} \big(\Tilde{\boldsymbol{\phi}}_{\mathrm{H},1}^* \Tilde{\boldsymbol{\phi}}_{\mathrm{H},2}^{\Ttran}  + \Tilde{\boldsymbol{\phi}}_{\mathrm{V},1}^* \Tilde{\boldsymbol{\phi}}_{\mathrm{V},2}^{\Ttran} \big)\hat{\vect{a}}_{\mathrm{H},2} =  \notag  \\
  &   - \hat{\vect{a}}_{\mathrm{H},1}^{\Htran} \vect{u}^*\vect{v}^{\Ttran} \otimes \boldsymbol{\phi}_{\mathrm{H}}^* \boldsymbol{\phi}_{\mathrm{V}}^{\Htran}\vect{E}_M   \hat{\vect{a}}_{\mathrm{H},2} + \hat{\vect{a}}_{\mathrm{H},1}^{\Htran}\vect{u}^*\vect{v}^{\Ttran} \otimes \boldsymbol{\phi}_{\mathrm{V}}^* \boldsymbol{\phi}_{\mathrm{H}}^{\Htran}\vect{E}_M  \hat{\vect{a}}_{\mathrm{H},2} \notag \\
  & = 0.
\end{align}
The PDAF of the expanded RIS thus equals the sum of the PDAFs of the two sub-RISs. Denoting the array response vector of the primary RIS for polarization H by $\hat{\vect{a}}^{\mathrm{P}}_{\mathrm{H}}$, we have $\hat{\vect{a}}_{\mathrm{H},1} = \hat{\vect{a}}_N(M\psi) \otimes \hat{\vect{a}}^{\mathrm{P}}_{\mathrm{H}}$, where $\hat{\vect{a}}_N(M\psi)$ is the equivalent array response vector for polarization H of an RIS having $2N$ elements with the inter-element spacing of $M\Delta_d$.  We also note that $\hat{\vect{a}}_{\mathrm{H},2} = e^{-j2MN\psi} \hat{\vect{a}}_{\mathrm{H},1}$. Hence, the PDAF reads as 
\begin{align}
   & \Tilde{A} (\varphi)= \notag \\ &\big| \vect{u}^{\Ttran} \hat{\vect{a}}_N(M\psi) \otimes \boldsymbol{\phi}_{\mathrm{H}}^{\Ttran}\hat{\vect{a}}^{\mathrm{P}}_{\mathrm{H}}\big|^2 + \big| \vect{v}^{\Ttran} \hat{\vect{a}}_N(M\psi) \otimes \boldsymbol{\phi}_{\mathrm{V}}^{\Htran}\vect{E}_M\hat{\vect{a}}^{\mathrm{P}}_{\mathrm{H}}\big|^2+ \notag \\
    & \big| \vect{u}^{\Ttran} \hat{\vect{a}}_N(M\psi) \otimes \boldsymbol{\phi}_{\mathrm{V}}^{\Ttran}\hat{\vect{a}}^{\mathrm{P}}_{\mathrm{H}}\big|^2 + \big| \vect{v}^{\Ttran} \hat{\vect{a}}_N(M\psi) \otimes \boldsymbol{\phi}_{\mathrm{H}}^{\Htran}\vect{E}_M\hat{\vect{a}}^{\mathrm{P}}_{\mathrm{H}}\big|^2 \overset{(a)}{=}\notag \\ 
    &  \big(| \vect{u}^{\Ttran} \hat{\vect{a}}_N(M\psi)|^2 + | \vect{v}^{\Ttran} \hat{\vect{a}}_N(M\psi)|^2 \big) \big( |\boldsymbol{\phi}_{\mathrm{H}} \hat{\vect{a}}^{\mathrm{P}}_{\mathrm{H}}|^2 + |\boldsymbol{\phi}_{\mathrm{V}} \hat{\vect{a}}^{\mathrm{P}}_{\mathrm{H}}|^2 \big) \overset{(b)}{=} \notag \\&2N A(\varphi),
\end{align}
where $(a)$ holds because $\vect{E}_M \hat{\vect{a}}^{\mathrm{P}}_{\mathrm{H}} = e^{-j2(M-1)\psi}\hat{\vect{a}}^{\mathrm{P*}}_{\mathrm{H}}$ and $|\boldsymbol{\phi}_i^{\Htran} \hat{\vect{a}}^{\mathrm{P*}}_{\mathrm{H}}| = |\boldsymbol{\phi}_i^{\Ttran} \hat{\vect{a}}^{\mathrm{P}}_{\mathrm{H}}|$, and $(b)$ is due to the fact that $\vect{u}$ and $\vect{v}$ are Golay complementary pairs of length $N$.

\bibliographystyle{IEEEtran}
\bibliography{IEEEabrv,refs}

\begin{thebibliography}{10}
\providecommand{\url}[1]{#1}
\csname url@samestyle\endcsname
\providecommand{\newblock}{\relax}
\providecommand{\bibinfo}[2]{#2}
\providecommand{\BIBentrySTDinterwordspacing}{\spaceskip=0pt\relax}
\providecommand{\BIBentryALTinterwordstretchfactor}{4}
\providecommand{\BIBentryALTinterwordspacing}{\spaceskip=\fontdimen2\font plus
\BIBentryALTinterwordstretchfactor\fontdimen3\font minus \fontdimen4\font\relax}
\providecommand{\BIBforeignlanguage}[2]{{%
\expandafter\ifx\csname l@#1\endcsname\relax
\typeout{** WARNING: IEEEtran.bst: No hyphenation pattern has been}%
\typeout{** loaded for the language `#1'. Using the pattern for}%
\typeout{** the default language instead.}%
\else
\language=\csname l@#1\endcsname
\fi
#2}}
\providecommand{\BIBdecl}{\relax}
\BIBdecl

\bibitem{Emil2019}
E.~Björnson, L.~Sanguinetti, H.~Wymeersch, J.~Hoydis, and T.~L. Marzetta, ``Massive {MIMO} is a reality—what is next? {Five} promising research directions for antenna arrays,'' \emph{Digit. Signal Process.}, vol.~94, pp. 3--20, 2019.

\bibitem{Guo2020}
H.~Guo, Y.-C. Liang, J.~Chen, and E.~G. Larsson, ``Weighted sum-rate maximization for reconfigurable intelligent surface aided wireless networks,'' \emph{IEEE Trans. Wireless Commun.}, vol.~19, no.~5, pp. 3064--3076, 2020.

\bibitem{Li2020}
Z.~Li, M.~Hua, Q.~Wang, and Q.~Song, ``Weighted sum-rate maximization for multi-{IRS} aided cooperative transmission,'' \emph{IEEE Wireless Commun. Lett.}, vol.~9, no.~10, pp. 1620--1624, 2020.

\bibitem{He2023BroadCoverage}
M.~He, J.~Xu, W.~Xu, H.~Shen, N.~Wang, and C.~Zhao, ``{RIS}-assisted quasi-static broad coverage for wideband mmwave massive {MIMO} systems,'' \emph{IEEE Transactions on Wireless Communications}, vol.~22, no.~4, pp. 2551--2565, 2023.

\bibitem{Li2021Golay}
F.~Li, Y.~Jiang, C.~Du, and X.~Wang, ``Construction of {Golay} complementary matrices and its applications to {MIMO} omnidirectional transmission,'' \emph{IEEE Trans. Signal Process.}, vol.~69, pp. 2100--2113, 2021.

\bibitem{Max2021}
M.~A. Girnyk and S.~O. Petersson, ``Efficient cell-specific beamforming for large antenna arrays,'' \emph{IEEE Trans. Commun.}, vol.~69, no.~12, pp. 8429--8442, 2021.

\bibitem{Max2022}
S.~O. Petersson and M.~A. Girnyk, ``Energy-efficient design of broad beams for massive {MIMO} systems,'' \emph{IEEE Trans. Veh. Technol.}, vol.~71, no.~11, pp. 11\,772 -- 11\,785, 2022.

\bibitem{ke2021linear}
J.~C. Ke \emph{et~al.}, ``Linear and nonlinear polarization syntheses and their programmable controls based on anisotropic time-domain digital coding metasurface,'' \emph{Small Struct.}, vol.~2, no.~1, p. 2000060, 2021.

\bibitem{Emil2022}
E.~Björnson, H.~Wymeersch, B.~Matthiesen, P.~Popovski, L.~Sanguinetti, and E.~de~Carvalho, ``Reconfigurable intelligent surfaces: A signal processing perspective with wireless applications,'' \emph{IEEE Signal Process. Mag.}, vol.~39, no.~2, pp. 135--158, 2022.

\bibitem{massivemimobook}
E.~Bj\"{o}rnson, J.~Hoydis, and L.~Sanguinetti, ``Massive {MIMO} networks: {Spectral}, energy, and hardware efficiency,'' \emph{Found. Trends Signal Process.}, vol.~11, no. 3-4, pp. 154--655, 2017.

\bibitem{Stein2000}
J.~Y. Stein, \emph{Digital Signal Processing—A Computer Science Perceptive}.\hskip 1em plus 0.5em minus 0.4em\relax New York: Wiley, 2000.

\bibitem{Wiener1930}
N.~Wiener, \emph{Generalized harmonic analysis}.\hskip 1em plus 0.5em minus 0.4em\relax Acta Math., 1930, vol.~55.

\bibitem{Golay1951}
M.~J.~E. Golay, ``Static multislit spectrometry and its application to the panoramic display of infrared spectra,'' \emph{J. Opt. Soc. Amer.}, vol.~41, no.~7, pp. 468--472, 1951.

\bibitem{3gpp}
{3GPP TR 36.873}, ``Study on {3D} channel model for {LTE},'' 3GPP, Tech. Rep., 2018.

\bibitem{tiwari2022ris}
K.~K. Tiwari and G.~Caire, ``{RIS}-based steerable beamforming antenna with near-field eigenmode feeder,'' \emph{arXiv preprint arXiv:2210.17239}, 2022.

\bibitem{Holzmann94}
W.~H. Holzmann and H.~Kharaghani, ``A computer search for complex {Golay} sequences,'' \emph{Australas. J. Comb.}, vol.~10, pp. 251--258, 1994.

\bibitem{Emil2020}
E.~Björnson, {\"O}.~Özdogan, and E.~G. Larsson, ``Intelligent reflecting surface versus decode-and-forward: How large surfaces are needed to beat relaying?'' \emph{IEEE Wireless Commun. Lett.}, vol.~9, no.~2, pp. 244--248, 2020.

\end{thebibliography}
\end{document}